\documentstyle[12pt]{article}
\begin{document}
\centerline{\large \bf Miron's Generalizations of}
\vskip4pt
\centerline{\large\bf  Lagrange and  Finsler Geometries:  }
\vskip4pt \centerline{\large \bf a Self--Consistent Approach to}
\vskip4pt
\centerline{\large\bf Locally Anisotropic Gravity}
\vskip10pt
 \centerline{\large \sf Sergiu I. Vacaru} \vskip8pt
{\small
\centerline{\noindent{\em Institute of Applied Physics, 
Academy of Sciences,}}
\centerline{\noindent{\em 5 Academy str., Chi\c sin\v au 2028,
Republic of Moldova}} \vskip5pt
\centerline{\noindent{ Fax: 011-3732-738149, E-mail: lises@cc.acad.md}}
 \centerline{--- --- --- --- ---}
\centerline{\noindent{\em Institute for Basic Research,}}
\centerline{\noindent{\em P. O. Box 1577, Palm Harbor, FL 34682, 
U. S. A.}}
\vskip3pt
\centerline{\noindent{ibr@gte.net,\ http://home1.gte.net/ibr}} }
\vskip10pt
\centerline{\large \it In honour of Academician Radu Miron\ 70th
birthday}
\vskip10pt

Modern gauge theories of high energy physics, investigations in
classical and quantum gravity  and recent unifications of superstring
theories (the so--called  $M$-- $F$-- and $S$--theories) are characterized
by a large application  of geometric and topological methods. There are 
elaborated a number of  Kaluza--Klein  models of space--time and proposed
different  variants of compactification of higher dimensions. One of
the still unsolved important physical problem is the definition of the
mechanism of (in general dynamical)  such compactification with a
rigorous "selection" of the four dimensional space--time physics from
the low energy dynamics of (super) string and supergravitational theories.
Another question of challenge of the modern physics is the local 
anisotropy of background radiation  and the development of a consistent
scenarious  of quantum and classical cosmology.

In our works \cite{v,vap,vjmp,vo} we have concerned the mentioned topics in
a more general context of modelling physical processes on (super)vector
bundles provided with nonlinear and distinguished connections and metric
structures (containing as particular cases both Kaluza--Klein spaces
and various extensions of Lagrange and Finsler spaces). We based our 
investigations on the fundamental  results of the famous R. Mirons's 
Romanian  school of Finsler geometry and its generalizations (as basic
references  we cite here some monographs and  recent works \cite{m}).

Perhaps, in the special literature there are cited more than a thousand
of works on Finsler geometry, its generalizations and applications. It 
is well known that a metric more  general than a Riemannian one was 
proposed in 1854 by B. Riemann and studied for the first time, 
in 1918, by P. Finsler who was a  post--graduate student of 
C. Carath\'eodory. The purpose of applications of  such generalized
metrics in thermodynamics and, more generally, in physics  was obvious.
At present time there are published tens of monographs containing 
Finsler--like physical theories. Nevertheless the bulk of physicists 
still persists on a broad implementation of Finsler geometry in modern
physics. This skepticism consists not only on a conservatism caused by
the predomination  of Riemann geometry (with some  extensions to 
Einstein--Cartan--Spaces with torsion and nonmetricity) or  by the 
"excessive complexity" of Finsler geometry for developing physical
theories. At first site the problem of construction of  a general 
"Finsler--physics" is very cumbersome and even unsurmountable. For
instance, the most fundamental  physical concepts of energy, momentum 
and rotation momentum are correspondingly strongly  related with the
supposed isotropy of space--time  with respect to time and space
translations and rotations (there is a group of automorphisms of the
flat Minkowski space, the so--called generalized  Lorentz, or  
Poincar\'e group, which also acts in the tangent bundle of the 
(pseudo)--Riemannian space). There are not  local symmetries on
spaces with generic local anisotropy and it "was concluded" that
not having even local (pseudo)--rotations  it impossible to construct a
consistent physical theory of Finsler--like space--time and to define
a theory of fundamental locally anisotropic field interactions because,
for example, electrons  are described by spinor fields, but spinors
are closely related with  the group of rotational isotropies which
could not be defined for  a general Finsler or Lagrange space.

It should be noted that the  development of physical theories in media
and  spaces with local anisotropy became a rather evident necessity
if there are rigorously analyzed a number of recent experimental data
from modern cosmology and astrophysics, which reflects substantial
anisotropies of the space--time structure at different state of
evolution of the Universe, and if there are studied various processes in
anisotropic, unordered and nonhomogeneous media, with dislocations and
disclinations and/or with fluctuations and anisotropic diffusion.
Supposing that  anisotropies are included only in the energy--momentum
tensors of matter (in order to develop some  cosmological scenarious),
which directs as the source in the Einstein equations the
dynamics of Universe, one does not obtain a self--consistent theory if
the condition of isotropy of metric (being a solution of field equations)
is imposed. The modern Cosmology requires a more general geometric
background with anisotropies of both matter and space--time metric.

In our works we tried to demonstrate that all basic difficulties connected
with the elaboration of {\bf locally anisotropic physics} are bridged over
if the problems are touched on  by modelling both generalized Finsler
geometries and physical theories on vector bundles provided with
compatible nonlinear and distinguished connections, metric, gauge and
spinor fields. Such idea and methods, being very important for a further
development of a number of directions in theoretical and mathematical
physics,  have been manifested in an explicit geometric form in the 
R. Miron and  co--authors works. In some initial and different forms the
idea  of modelling of  physical interactions on spaces with different 
geometric  structures was contained in the Yano and Ishihara monograph
 \cite{YI} (lifts of geometric object on the total space of tangent
bundles) and in the  A. Z. Petrov \cite{Petrov} and N. S. Sinyukov
\cite{Sinyukov}  works (deformation of connections and  metric structures
by maps generalizing conformal transforms).

The first our contributions to extensions of  Finsler geometry and
applications  were  the formulation of the theory of nearly autoparallel
maps of generalized Lagrange and Finsler spaces \cite{vo} and the
proposal of definition of conservation laws for locally anisotropic
field interactions by using such generalizations of conformal maps.
The next step was the differential geometry of Clifford and spinor
fibrations and of affine and linear frame bundles provided with
nonlinear connection structure \cite{vjmp} and in consequence the
formulation  of the theory of locally anisotropic field interactions.

In order to demonstrate that the locally anisotropic physics is naturally
contained in modern (super) string theories we investigated the low
energy limit of string locally anisotropic perturbations and proved
that there are alternative variants when the gravitational and matter
field interactions could be locally isotropic or anisotropic \cite{vap}. 
We formulated the supergeometry of vector superbundles enabled with
general nonlinear connections and extended the diagram methods of
superstrings to the case of higher order anisotropies. Another directions
of our investigations is connected with higher order supersymmetric
stochastic processes and superdiffusion. The basic results are summarized
in monographs \cite{v}.

The main conclusion of this note is that after the R. Miron
and co--authors idea of modelling  both geometric constructions and
physical theories was applied in our works  in a manner as to unify
both Kaluza--Klein and generalized Finsler (super)spaces by imbedding
them in modern  string theories  it was  compelled that  the
present--day physics can be generalized as to include locally 
anisotropic  interactions and spaces.


{\small
\begin{thebibliography}{99}
\bibitem{m}
 R. Miron, {\it The Geometry of Higher Order Lagrange Spaces: Applications
 to Mechanics and Physics} (Kluwer Academic Publishers, Dordrecht,
 Boston, London, 1997);\
R. Miron, {\it The Geometry of Higher Order Finsler Spaces} (Hadronic
 Press, Instituto per la Ricerca di Base, Italy, 1997);\
 R. Miron and M. Anastasiei, {\it Vector Bundles. Lagrange
Spaces. Application in Relativity} (Academiei, Romania, 1987) [in Romanian],
 English translation: (Geometry Balkan Press, 1996)\
  R. Miron and M. Anastasiei, {\it The Geometry of Lagrange
Spaces: Theory and Applications} (Kluwer Academic Publishers, Dordrecht,
Boston, London, 1994);\
R. Miron and Gh. Atanasiu, {\sl Revue Roumaine de Mathematiques Pures
 et Appliquees} {\bf XLI} $N^{os}$ 3--4 (1996) 205; 237; 251
\bibitem{Petrov}
 A. Z. Petrov, {\it Modelling of physical fields,}
 in {\sl Gravitation and the Theory of Relativity,} ed. A. Z. Petrov,
 N 4 \& 5 (Kazani University Press, URSS, 1968) 7;\
 A. Z. Petrov, {\sl Doklady Academii Nauk SSSR} {\bf 186} (1969) 1302;\
 {\bf 190} (1970) 305;\ A. Z. Petrov, {\it On Modelling of Gravitational
 Fields.} Preprint ITF-71-14R, Kiev, 1971 [in Russian]
\bibitem{Sinyukov}
 N. S. Sinyukov, {\it Geodesic Maps of Riemannian Spaces}
 (Nauka, Moskow, 1979) [in Russian]
\bibitem{v}
 S. Vacaru, {\it Interactions, Strings and Isotopies in Higher Order
 Anisotropic Superspaces} (Hadronic Press, Palm
 Harbor, FL, USA, 1998);\
  S. Vacaru, {\it Field Interactions, Strings, and Diffusion in
 Local\-ly An\-isotropic (Super)Spaces} (under consideration);\
 E--prints: hep--th/9611091; physics/9705030; physics/9706038;
 gr--qc/9604014
\bibitem{vap}  S. Vacaru, {\it Locally anisotropic gravity and strings,}
 {\sl Ann. Phys (N. Y.),} {\bf 256} (1997) 39;\
   S. Vacaru, {\it Superstrings in Higher Order Extensions
of Finsler Superspaces}  {\sl Nucl. Phys. B}, {\bf 434} (1997) 590;\
E--print: hep--th/9611034
\bibitem{vjmp}  S. Vacaru, {\it Spinor structures and nonlinear connections
in vector bundles, generalized Lagrange and Finsler spaces,} {\sl J. Math.
Phys} {\bf 37} (1996) 508;\
  S. Vacaru and Yu. Goncharenko, {\it Yang-Mills fields and
gauge gravity on generalized Lagrange and Finsler spaces,} {\sl Int. J.
Theor. Phys.} {\bf 34} (1995) 1955;\
 E--print: dg--ga/9609004;
\bibitem{vo}
  S. Vacaru and S. Ostaf, {\it Nearly autoparallel maps of
Lagrange and Finsler spaces,} in {\sl Lagrange and Finsler Geometry,} eds.
P. L. Antonelli and R. Miron, (Kluwer Academic Publishers, Dordrecht,
Boston, London, 1996) 241;\
 S. Vacaru and S. Ostaf, {\it Twistors and nearly
autoparallel maps,} {\sl Rep. Math. Phys.} {\bf 37} (1996) 309; E-print:\
gr-qc/9602010
\bibitem{YI} K. Yano and S. I. Ishihara, {\it Tangent and Cotangent
 Bundles. Differential Geometry} (Marcel Dekker, New York, 1973)
\end{thebibliography}
}
\end{document}